\documentclass[twocolumn,aps,reprint,numerical,superscriptaddress]{revtex4}

\usepackage[T1]{fontenc}
\usepackage[utf8]{luainputenc}
\usepackage{amsmath,amssymb}
\usepackage{graphicx}
\usepackage[english]{babel}

\newcommand{\ie}{{\it i.e.}}

\newcommand{\celegans}{\textit{C.elegans}}
\newcommand{\hr}{Hindmarsh-Rose}

\begin{document}

\title{Chimera-like states in modular neural networks}
\author{Johanne Hizanidis} 
\affiliation{Crete Center for Quantum Complexity and Nanotechnology, Physics Department, University of Crete, 71003 Heraklion, Greece} 
\affiliation{National Center for Scientific Research ``Demokritos'', 15310 Athens, Greece}

\author{Nikos E. Kouvaris} 
\affiliation{Department of Physics, University of Barcelona, Mart\'i i Franqu$\grave{e}$s 1,E-08028, Barcelona, Spain}

\author{Gorka Zamora-L\'opez} 
\affiliation{Center for Brain and Cognition, Universitat Pompeu Fabra, Barcelona, Spain}
\affiliation{Department of Information and Communication Technologies, Universitat Pompeu Fabra, Barcelona, Spain}

\author{Albert D\'iaz-Guilera} 
\affiliation{Department of Physics, University of Barcelona, Mart\'i i Franqu$\grave{e}$s 1,E-08028, Barcelona, Spain}

\author{Chris G. Antonopoulos} 
\affiliation{Department of Mathematical Sciences, University of Essex, Wivenhoe Park, CO4 3SQ Colchester, UK}


\begin{abstract}
Chimera states, namely the coexistence of coherent and incoherent behavior, were previously analyzed in complex networks. However, they have not been extensively studied in modular networks. Here, we consider a neural network inspired by the connectome of the \celegans{} soil worm, organized into six interconnected communities, where neurons obey chaotic bursting dynamics. Neurons are assumed to be connected with electrical synapses within their communities and with chemical synapses across them. As our numerical simulations reveal, the coaction of these two types of coupling can shape the dynamics in such a way that chimera-like states can happen. They consist of a fraction of synchronized neurons which belong to the larger communities, and a fraction of desynchronized neurons which are part of smaller communities. In addition to the Kuramoto order parameter $\rho$, we also employ other measures of coherence, such as the chimera-like $\chi$ and metastability $\lambda$ indices, which quantify the degree of 
synchronization among communities and along time, respectively. We perform the same analysis for networks that share common features with the \celegans{} neural network. Similar results suggest that under certain assumptions, chimera-like states are prominent phenomena in modular networks, and might provide insight for the behavior of more complex modular networks.
\end{abstract}

\maketitle

\section*{Introduction}

One of the most challenging complex system is the human brain in which neurons and their interconnections through synapses form a very complicated structure, the cortical network. The complexity of the circuitry of the nervous system of the human brain is still a big challenge to be resolved as it contains about 86 billion neurons and thousands times more synapses \cite{Azevedoetal2009}. Neurons are linked  together to perform certain tasks and cognitive functions, such as pattern recognition, function approximation, data processing, etc. It has been revealed that the cortical network is hierarchical and clustered with a complex connectivity \cite{Hilgetag_2004}, known as the modular organization of the brain. This demands an inherent parallel nature of brain computations \cite{Meunieretal2010}. Modular processors have to be sufficiently isolated and dynamically differentiated to achieve independent computations, but at the same time also globally connected to be integrated in coherent functions \cite{
Zamora2010,Meunieretal2010}. A possible network description for this modular organization is that brain networks may be small-world structured \cite{Heetal2007} with properties similar to many other complex networks \cite{Stam2004}. This viewpoint has been driven by the systematic finding of small-world topology in a wide range of human brain networks derived from structural \cite{Heetal2007} and functional \cite{Eguiluzetal2005} studies. Small-world topology has also been identified in functional cortical neural networks in mammals \cite{Yuetal2008} and also in the nervous system of the nematode \textit{Caenorhabditis elegans} (\celegans) soil worm \cite{Wattsetal1998,Antonopoulosetal2015}. This topology seems to be relevant for brain function, as it is affected by diseases \cite{Yuetal2007}, normal ageing, and by pharmacological blockade of dopamine neurotransmission \cite{Achardetal2007}.

In recent years, enormous research has been devoted on the \celegans~which has revealed its ability to learn about mechano-, chemo- and thermo-sensory stimuli \cite{Bargmann2006,Kapla1993}. It was also shown that its neural system has the ability to distinguish between tastes, odours or any indication related to the presence or absence of food. Moreover, it shows different kinds of learning behavior, including associative learning such as classical conditioning and differential classical conditioning, and non-associative forms of learning such as habituation and dishabituation \cite{Gallyetal2003}. These properties, though quite ``simple'', are reminiscent of the human brain ability to adapt to different stimuli and environments. Moreover, many neurotransmitters such as Glutamate, GABA, Acetylcholine and Dopamine are common in the human brain and the \celegans~neural network. The genome of the \celegans~is almost 30 times smaller than that of humans but still, encodes almost 22000 proteins and it is almost 
35\% similar to that of humans \cite{Blumenthaletal1996}. Therefore, the study of the \celegans~neural network may give an insight of the possible behavior of more complex systems, such as the human brain.

Neural networks, among other complex systems, self-organize in ways that synchronous spatiotemporal patterns can emerge. Insightful findings regarding synchronization in complex networks have been reviewed in Ref. \cite{Arenasetal2008} and, recently, synchronization in complex modular or clustered networks has been investigated in Ref. \cite{Gardenesetal2011}. It appears as the interplay between the intrinsic dynamics associated to the nodes of the network and, its topology and connecting functions \cite{Nicosia2013}. Synchronization in neural networks is important for normal and various cognitive functions \cite{Mathias2011}, but may also reflect pathological brain states \cite{Traub_1982}. There is also increasing evidence that various types of brain diseases such as Alzheimer's disease, schizophrenia and brain tumors may be associated with deviations of the network topology from the optimal small-world pattern \cite{Stametal2007}. It has been found that burst synchronization of neural systems may be 
strongly influenced by many factors, such as coupling strengths and types \cite{Belykh_2005}, noise \cite{Buric_2007}, and the existence of clusters in neural networks.

Over a decade ago, a very peculiar phenomenon of synchronization was reported in coupled oscillators, where a hybrid state combining both coherent and incoherent parts can spontaneously emerge \cite{PhysRevA.39.4835,KUR02a}, which was later termed chimera state \cite{ABR04}. Surprisingly, these states were first found in systems of identical oscillators coupled  with a symmetric interaction function. Since then, there has been increasing interest in chimera states \cite{LAI09,MOT10,MAR10,OME10a,BOU14,LAZ15} and it has been shown that they are not limited to phase oscillators, but can also appear in a large variety of different systems including neural systems \cite{OME13,HIZ13,OME15} which is the focus of the present work. Apart from the classical chimera states, which consist of a coherent and incoherent domain, chimeras with multiple incoherent regions \cite{SET08,OME13,VUE14a}, as well as amplitude chimera states \cite{SET14} and chimera death states \cite{ZAK14} have also been recently reported. The 
study of chimera states goes beyond numerical observations. Experimental verification was first demonstrated in chemical \cite{TIN12} and optical \cite{HAG12} systems. Further experiments were realized in mechanical \cite{MAR13}, electronic \cite{LAR13}, and electrochemical \cite{SCH14a, WIC13} oscillator systems. Until recently, it was widely assumed that identical elements and symmetric coupling topology were prerequisites for chimera states. These limitations, however, can be overcome and chimera-like states can also be found in systems with non-identical elements, or with non-regular or even global topologies (see \cite{PAN15} and references within). Potential applications of chimera states in nature include bump states in neural systems \cite{LAI01,SAK06a} and the phenomenon of unihemispheric sleep in birds and dolphins \cite{RAT00}, which sleep with one eye open, meaning that half of the brain is synchronized with the other half being desynchronized. This is relevant for studies of synchronization in 
community-based networks where only a few works have focused on this interesting phenomenon \cite{SHA10,WIL12d}.

Our aim is to contribute in this direction by considering a topology based on the \celegans~\cite{cmtkdataset} network, whose neurons are found, by employing a community detection algorithm \cite{Ponsetal2005}, to be organized in six communities (see Fig.~\ref{fig1}). We assume that neurons are connected using two types of synapses: electrical and chemical \cite{footnote1}, which connect the neurons within and across the communities, respectively~\cite{Baptistaetal2010,Antonopoulosetal2015}.

We model the neuron dynamics in terms of the \hr~system. Our primary focus lies on the conditions for the existence of chimera-like states in this modular neural network, and at the same time on the relationship between the properties of these states with the topological characteristics of the considered network.

\section*{Results}

We first study the level of synchronization within each community and of the entire network as a function of the electrical and chemical couplings. To measure synchronization, we use the Kuramoto order parameter $\rho$ (Eq.~\eqref{z_t}). The parameter $\rho$ is bounded in the interval $[0,1]$ and is equal to 1 when neurons in the considered population are completely synchronized and 0 when they are totally desynchronized. Note that, one actually averages $\rho(t)$ over time to obtain the value to which the order parameter converges in a sufficiently long time interval.

Figures~\ref{fig2}(a-g) show the order parameter for each community and for the entire network, in the $(g_{ch},g_{el})$ parameter space. All six communities have a region of high synchronization for low chemical and high electrical coupling, which is also reflected in the parameter space of the entire network (Fig.~\ref{fig2}(g)). This is reasonable since in this case, the neurons are strongly connected within each community and at the same time the coupling between communities is weak, meaning that the different populations do not affect each other significantly. In particular, for communities 2 and 4, this region of synchronization extends also to smaller electrical couplings. This is because these communities have the largest number of nodes and electrical links within them (see Fig.~\ref{fig1}(a) and \ref{fig1}(b)). Small communities are not easily synchronizable due to the strong influence from the large ones and to the sparsity of their electrical synapses.

Apart from this region which is common in all communities, there are other ``islands'' of synchronization where one or more communities achieve high values of the order parameter. Prominent examples are Fig.~\ref{fig2}(c) and especially Fig.~\ref{fig2}(f), where communities 3 and 6 show high synchronization levels for high chemical couplings with the rest of them and the network globally, being incoherent. For community 6, in particular, this region is located at high values of the chemical coupling and gradually shrinks as the electrical coupling increases. This means that community 6 is dominated by the connected communities, a fact which is also clear from the size of the arrowheads directing to it in Fig.~\ref{fig1}(b). This effect, although weaker, is also responsible for the additional high synchronization area in the parameter space of community 3, located at large values of both chemical and electrical couplings.

Another measure related to synchronization, which is very relevant in modular networks, is the metastability index $\lambda$, which measures the coherence among the communities along time. Figure~\ref{fig2}(h) shows the parameter space for $\lambda$ (Eq.~\eqref{lambda}). It is clear that the region in which $\lambda$ attains values close to zero coincides with the region where all communities are in a synchronized state. In other parts of the parameter space $\lambda$ achieves higher values, indicating that the system often switches between synchronous and asynchronous states.

Figure~\ref{fig2}(i) corresponds to the parameter space for the chimera-like index (Eq.~\eqref{chi}). It is evident that $\chi$ achieves its highest values in the two synchronization ``islands'' of communities 3 and 6, as well as on the border of the common synchronization region separating coherent from incoherent behavior. All of the aforementioned parameter spaces allow us to gain a complete picture of the collective behavior in our system, both in the community and global scale. 

In order to highlight some characteristic patterns that emerge in our system, we select 3 points of interest on the $(g_{ch},g_{el})$ parameter space, marked by letters {\bf A}, {\bf B} and {\bf C}. They are chosen so that the following three cases are covered, {\bf A}: both $\lambda$ and $\chi$ are low-valued, {\bf B}: metastability prevails, \ie{} $\lambda \gg \chi$ (when normalized), and {\bf C}: ``chimera-like'' states prevail over metastability \ie{} $\chi\gg\lambda$ (when normalized). Figure~\ref{fig3} shows the space-time plots of the membrane potential $p$ (see Eqs.~\eqref{HR_model_Nneurons}) for the three points, one typical snapshot in time where each community is coloured in accordance to Fig.~\ref{fig1}, and the time series of node 100, which belongs to community 3. The nodes in the communities are relabelled so that each community is placed next to each other.

Point {\bf A} corresponds to low metastability and low chimera-like index (see Fig.~\ref{fig2}(h) and \ref{fig2}(i)). This means that the network as a whole does not switch in time to different synchronization patterns frequently, and simultaneously, the 6 communities are, to a large extent, in synchrony with each other (see also Supplementary Movie S1). This is expected for the combination of high electrical and low chemical coupling and, is in agreement with the high value of the global order parameter shown in Fig.~\ref{fig2}(g). The corresponding time series exhibits spiking behavior with short quiescent periods between spike appearance (Fig.~\ref{fig3}(a)).

The metastability effect for low chimera-like index is shown for point {\bf B}. In the space-time plot of Fig.~\ref{fig3}(b) this is illustrated by the rather regular spatial pattern (due to low $\chi$), which alternates in time between slow quiescent periods (yellow-red) and fast spiking intervals (blue-green) that correspond to synchronous and incoherent behaviors, respectively (see also Supplementary Movie S2). From the time series, it is evident that, for these parameters, the system is in the bursting regime.

Point {\bf C} corresponds to a chimera-like state, in which the metastability of the system attains low values. Communities 2 and 4 are (on average) the most synchronized ones, illustrated in Fig.~\ref{fig3}(c) and also verified by the corresponding high values of the order parameter (Fig.~\ref{fig2}(b), (d)). This is reasonable, since these two communities are the largest ones in the network (see Fig.~\ref{fig1}(a), (b)). The remaining communities alternate more frequently between spiking and quiescent behavior (see also Supplementary Movie S3). They are also perturbed by the inputs from the large communities making it harder for them to synchronize, as is also captured in the snapshot of Fig.~\ref{fig3}(c). 

The above analysis has been carried out for other modular networks (see Supplementary Information) and the results are in qualitative agreement with those presented in this section for the \celegans~-based neural network. This agreement shows that our results do not depend on the community detection algorithm but, rather, on the interaction of the dynamics with the topology of the considered network.

\section*{Discussion}

In this work we have quantified and compared certain  measures of dynamical complexity, such as synchronization, which may underpin the ``differentiation'' in sub-domains and the ``integration'' as the system exhibits coherent behavior as a whole \cite{Srinivasanetal1999} and, the metastability $\lambda$ and chimera-like $\chi$ indices that allow for the quantification of the degree of metastability and chimera-like behavior exhibited by the system and its communities. In \cite{Antonopoulosetal2015}, various statistical quantities associated with the \textit{C.elegans} neural network, such as the global clustering coefficient, the average of local clustering coefficients, the mean shortest path, the degree probability distribution function of the network and the small-worldness measure have been computed. The latter property is characterized by a relatively short minimum path length on average between all pairs of nodes, together with a high clustering coefficient. Even though small-worldness captures 
important aspects of complex networks at the local and global scale of the structure, it does not provide information about the intermediate scale. These properties can be better described by the community structure or modularity of the network. Since nodes within the same module are densely intra-connected, the number of triangles in a modular network is larger than in a random graph of the same size and degree distribution, while the existence of a few links between nodes in different modules plays the role of topological shortcuts in the small-world topology. Networks characterized by this property tend to be small-world, with a high clustering coefficient and short path length with respect to random networks.

Here, we modeled a neural network based on the \celegans{} soil worm connectome in terms of \hr{} dynamics and divided its network into six communities employing the walktrap method. Based on the numerical simulations of this system, we analyzed conditions under which coherent and incoherent neural phase synchronization emerges simultaneously among communities for certain values of the chemical and electrical couplings. We related this phenomenon to structural network characteristics, such as the number of nodes of the communities, the ratio of the number of chemical synapses connecting their nodes, the ratio of the number of chemical synapses divided by the mean degree of the target community, the absolute participation and contribution of nodes to the modular structure and finally, to participation and global hubness. We found out that none of the nodes with low participation are hubs of the network, which is characteristic  of networks with well-defined communities, even though we found a significant 
number of nodes with intermediate participation, showing that most of the neurons share connections across several communities. We also found that the hubs of the network are among the nodes with the largest participation, extending their connections among most of the communities. All these findings clearly delineate the influence of one community on the others, and especially that of the two largest ones.
 
The phenomenon of chimera-like states we identified here is reminiscent of classical chimeras observed in the Kuramoto and other models. In our study, we showed that chimera-like states are spontaneously formed at chemical and electrical coupling values for which the chimera-like index is much higher than the metastability index and that they are driven by the largest communities, which were found to be the most influential ones. This remarkable behavior was found to be prominent in the two synchronization ``islands'' of two communities, as well as on the border of the common synchronization region separating coherent from incoherent behavior. In the Supplementary Information, we compared our results with those for modular networks sharing common features with that of the \celegans{}-based network, grouped into six Erd\H os-R\'enyi and small-world communities, respectively. We concluded that chimera-like states can also be found in the modular networks, suggesting that they appear to be prominent phenomena 
which 
might provide insights for the functioning of the human brain.


\section*{Methods}

\paragraph*{\bf The Hindmarsh-Rose System.}

\begin{figure*}[ht!]
\begin{center}
\includegraphics{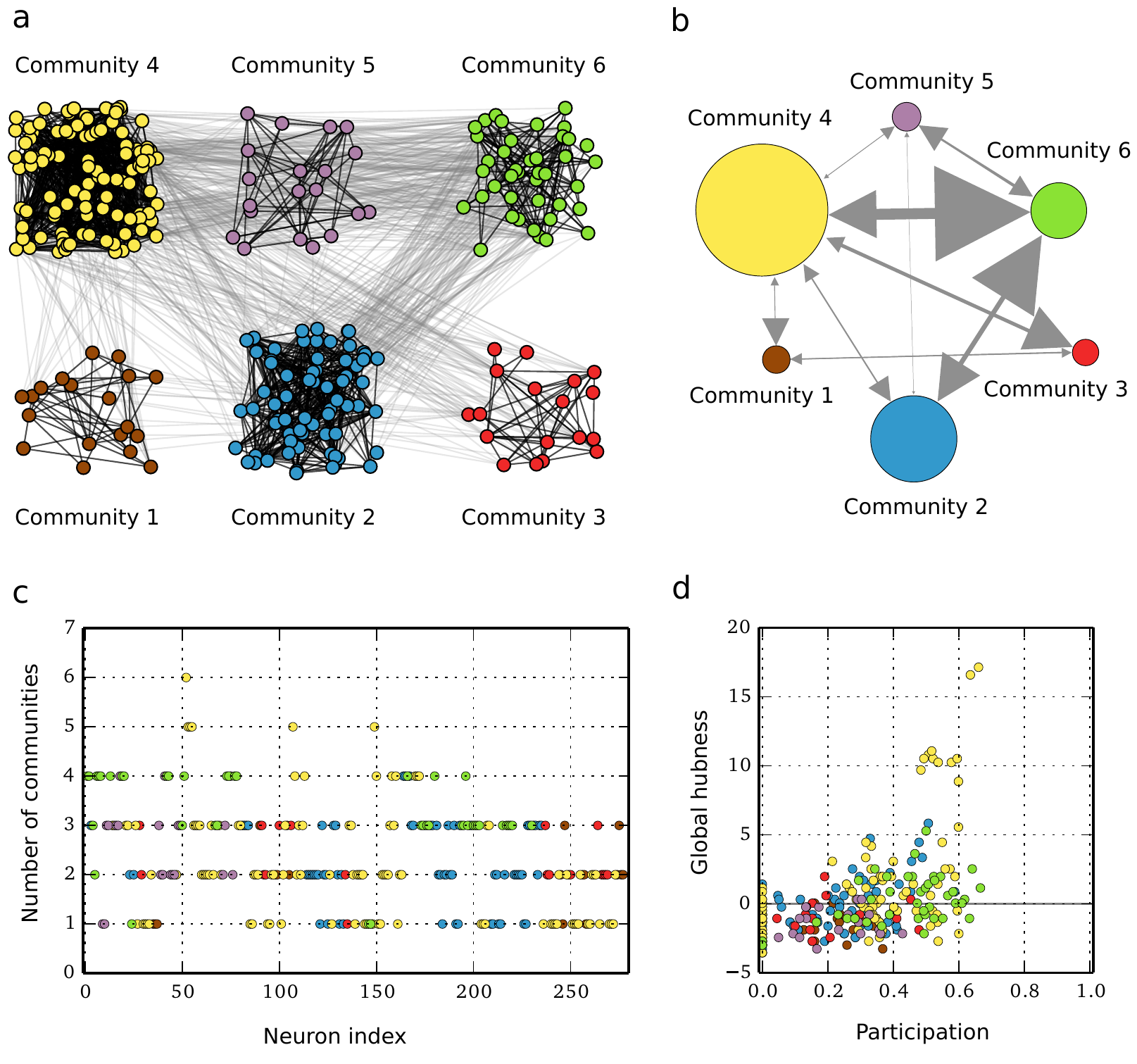}
\end{center}
\caption{{\bf Modular organization of the \celegans~neural network and contribution of neurons to the modular structure.} (a) Employing the walktrap method \cite{Ponsetal2005}, six communities have been detected in the neural network of \celegans{}. Different colors are assigned to neurons of different communities, 1: brown (19 neurons), 2: blue (69 neurons), 3: red (18 neurons), 4: yellow (108 neurons), 5: purple (20 neurons), 6: green (43 neurons), and are used throughout all figures. Different type of synapses are depicted by links of different color; black links denote electrical coupling between neurons within the same community, whereas gray links represent chemical coupling between neurons across different communities. (b) The communities are represented by circles of different size which is proportional to the number of their nodes. Directed links connecting the communities clearly delineate the influence of one community to the other. Their width is proportional to the number of chemical synapses 
connecting their nodes and the size of their arrows is proportional to the number of those chemical synapses divided by the mean degree of the target community. (c) Absolute participation and contribution of nodes to the modular structure. (d) Relation between participation and global hubness. None of the nodes with low participation is a hub of the network, what is characteristic of networks with well-defined communities. However, we find a significant number of nodes with intermediate participation, showing that most neurons share connections across several communities. The hubs of the network are among the nodes with the largest participation, extending their connections among most of the communities.}
\label{fig1}
\end{figure*}

\begin{figure*}[ht!]
\begin{center}
\includegraphics{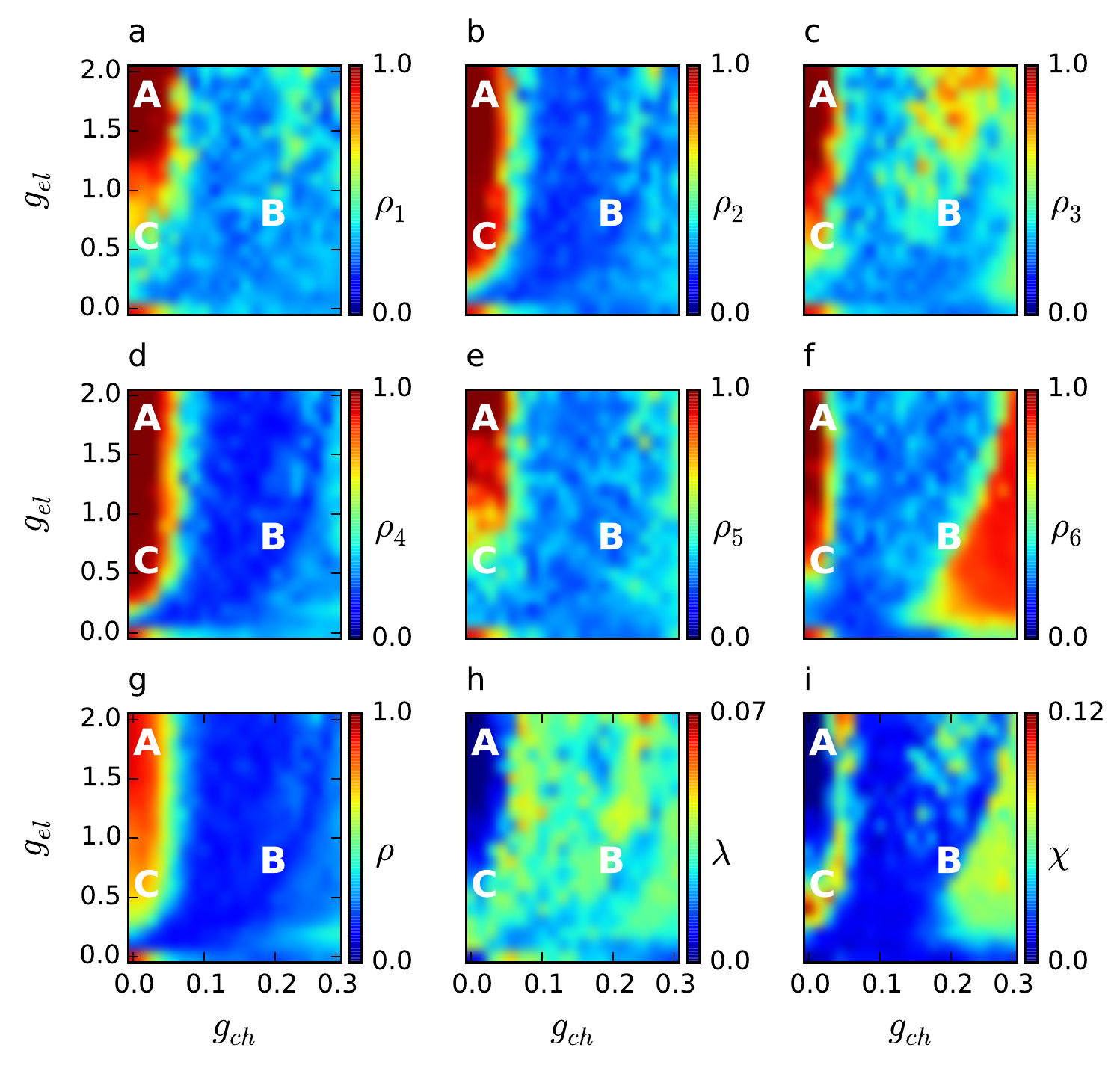}
\end{center}
\caption{{\bf Parameter spaces.} Density plots show the order parameters as well as the metastability and the chimera-like index in the same $(g_{ch},g_{el})$ plain. The order parameter of each community $\rho_{1,\ldots,6}$ is shown in (a)-(f), and of the entire network in (g). The metastability index is shown in (h) and the chimera-like index in (i) (note the different range of values in the color bar in (h) and (i)). The marked points {\bf A} ($g_{ch}=0.015$, $g_{el}=1.7$), {\bf B} ($g_{ch}=0.18$, $g_{el}=0.7$) and {\bf C} ($g_{ch}=0.015$, $g_{el}=0.5$) denote three different dynamical regimes. Synchronization is clear within each community in the regime where point {\bf A} is chosen and corresponds to low values of both, $\lambda$ and $\chi$. Point {\bf B} characterizes a regime where the synchronization index of each community and of the entire network is very low, $\chi$ is also low and $\lambda$ is high. This indicates a highly metastable behavior and desynchronous dynamics. An interesting behavior is 
found 
on the edge that separates those two regimes, where point {\bf C} is located. Some communities (obviously 2 and 4, less clearly 6) are synchronous and others (1, 3 and 5) are desynchronous. Furthermore, $\lambda$ is low, whereas $\chi$ is high, meaning that this state is not metastable and chimera-like dynamics can be observed. The dynamical behavior in these points is illustrated clearly in Fig.~\ref{fig3}.}
\label{fig2}
\end{figure*}

\begin{figure*}[ht!]
\begin{center}
\includegraphics{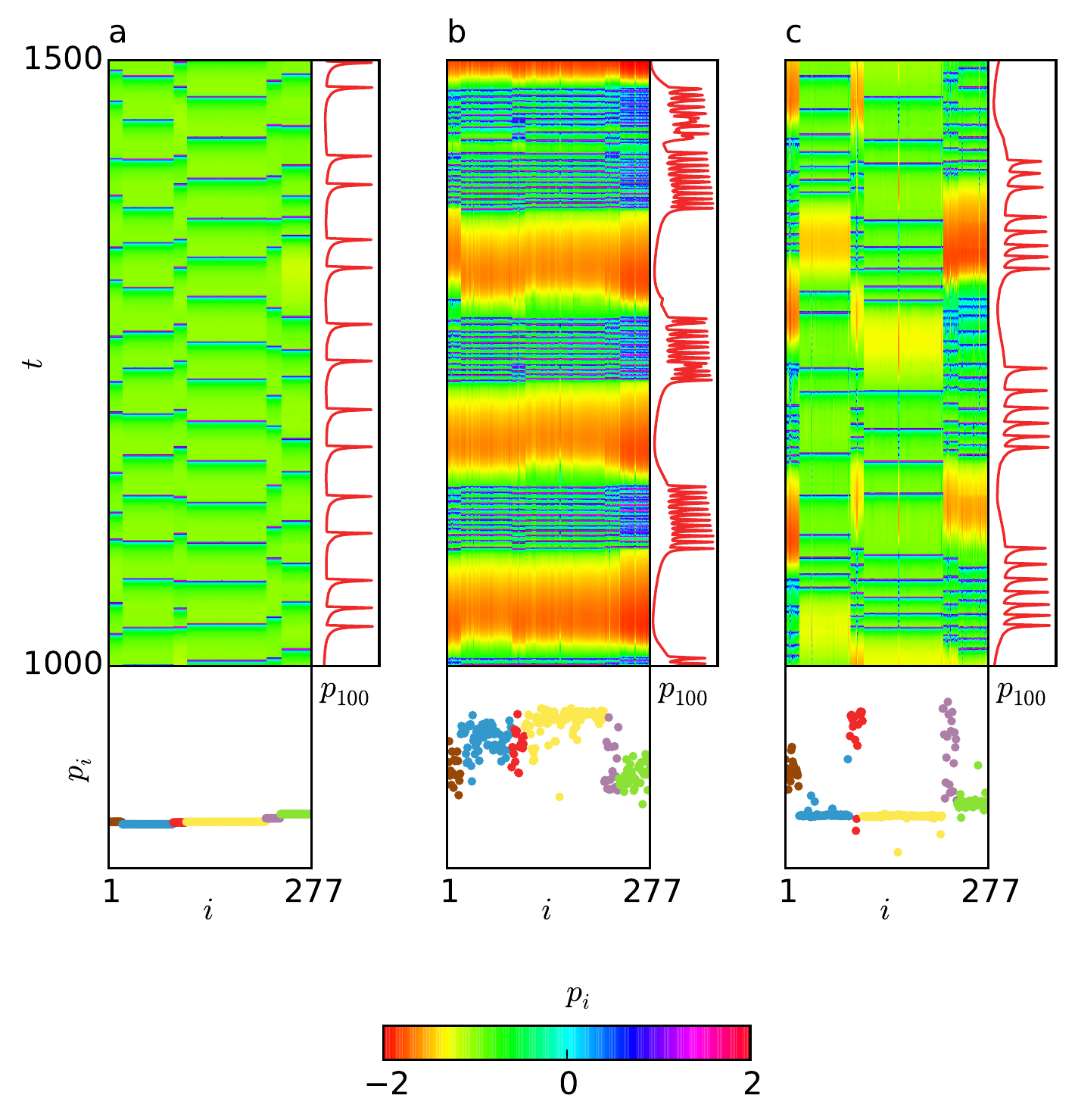}
\end{center}
\caption{{\bf Synchronous, metastable and chimera-like states.} (a) The spatiotemporal evolution of $p_i$ (upper left), with a time series of the neuron with index 100 of community 3 (upper right) and a snapshot of the system state (bottom) are shown for $g_{ch}=0.015$ and $g_{el}=1.7$, which correspond to point {\bf A} of Fig.~\ref{fig2}. Both, $\lambda$ and $\chi$ are very low, thus synchronization is clearly shown. (b) The same plot for point {\bf B} ($g_{ch}=0.18$, $g_{el}=0.7$). Here, the metastable index $\lambda$ is high, while the chimera-like index $\chi$ is low, thus a desynchronized state is depicted. (c) The same plot for point {\bf C} ($g_{ch}=0.015$, $g_{el}=0.5$), where the metastable index $\lambda$ is low and the chimera-like index $\chi$ is high. A chimera-like state is illustrated here, where neurons in communities 1 (brown), 3 (red), 5 (purple) and 6 (green) are desynchronized, whereas neurons in communities 2 (blue) and 4 (yellow) are synchronized. Neurons are ordered according to their 
community. Same color code as in Fig.~\ref{fig1} is used in the snapshots.}
\label{fig3}
\end{figure*}

In the current study we aim to analyze how neural dynamics can be collectively shaped by the coaction of two distinct types of synapses, namely of electrical and chemical. For this purpose, we consider a \celegans-based neural network \cite{Varshneyetal2011}, where we employ a community detection method that finds six communities. A schematic representation of this network is shown in Fig.~\ref{fig1}(a). Based on the detected communities, we assume that neurons within the same community are connected with electrical synapses (black links in Fig.~\ref{fig1}(a)) and neurons across communities with chemical synapses (gray links in Fig.~\ref{fig1}(a)). Finally, we endow each neuron with \hr{} dynamics.

Before we incorporate in the dynamics the coupling terms arising from the electrical and chemical synapses of the individual neurons, we briefly discuss the different characteristics and functionality of these types of synapses. A synapse is a junction between two neurons and serves as the means by which neurons communicate with each other. In particular, an electrical synapse (electrical link) is a physical connection between two neurons that allows electrons to pass through neurons by a very small gap between them. Electrical synapses are bidirectional and of a local character, happening between neurons which are spatially very close. Mutual coupling through these synapses promotes phase synchronization and coherence, resulting into groups of synchronized neurons. On the other hand, chemical synapses (chemical links) are typically unidirectional and the pre-synaptic signals are transmitted via release of neurotransmitters from the pre-synaptic neuron, which attaches to receptors at the post-synaptic neuron.
 Depending on the neurotransmitter, a chemical synapse can either be excitatory or inhibitory. Since the empirical methods used to identify the neuronal wiring in the \celegans~cannot distinguish between the two types of synapses, here we consider only excitatory chemical ones \cite{Varshneyetal2011}. 

By applying the \hr~dynamics to the nodes of the network that incorporates both types of synapses, we create an undirected neural network, in which neurons are connected by electrical (linear diffusive coupling) and chemical (nonlinear coupling) synapses, described by the equations,
\begin{eqnarray}
\label{HR_model_Nneurons}
\dot{p}_i&=&q_i-a p_i^3+bp_i^2-n_i+I_{\text{ext}}\nonumber\\
 &+&g_{el}\sum_{j=1}^{N}L_{ij}H(p_j)
 -g_{ch}(p_i-V_{\text{syn}})\sum_{j=1}^{N}T_{ij}S(p_j),\nonumber\\
\dot{q}_i&=&c-dp_i^2-q_i,\nonumber\\
\dot{n}_i&=&r[s(p_i-p_0)-n_i],
\end{eqnarray}
where $i=1,\ldots,N$ is the neuron index, $p_i$ is the membrane potential of the $i$-th neuron, $q_i$ is associated with the fast current, either $Na^{+}$ or $K^{+}$, and $n_i$ with the slow current, for example $Ca^{2+}$. The parameters of Eqs. \eqref{HR_model_Nneurons} are chosen such that $a=1$, $b=3$, $c=1$, $d=5$, $s=4$, $p_0=-1.6$ and $I_{\text{ext}}=3.25$, for which the system exhibits a multi-scale chaotic behavior characterized as spike bursting. $r$ modulates the slow dynamics of the system and is set to $0.005$ so that each neuron lies in the chaotic regime. For these parameters, the \hr~model enables the spiking-bursting behavior of the membrane potential observed in experiments made with single neurons \textit{in vitro}. It is a relatively simple model that provides a good qualitative description of many different patterns empirically observed in neural activity.  The connectivity structure of the electrical synapses is described in terms of the Laplacian matrix $\mathbf{L}$, whose elements are 
defined as $L_{ij}=E_{ij}-\delta_{ij}k_i$, where $\delta_{ij}=1$ if $i=j$, and $\delta_{ij}=0$ otherwise. $\mathbf{E}$ is an adjacency matrix whose elements are  $E_{ij}=1$ if there is an electrical synapse connecting the neurons $i$ and $j$, and $E_{ij}=0$ otherwise. The strength of the electrical coupling is given by the parameter $g_{el}$ and its functionality is governed by the linear function $H(p)=p$. 

The connectivity structure of the chemical synapses is described in terms of the adjacency matrix $\mathbf{T}$, whose elements are $T_{ij}=1$ if there is a chemical synapse between neurons $i$ and $j$, and $T_{ij}=0$ otherwise. The chemical coupling is nonlinear and its functionality is described by the sigmoidal function $S(p)=\{1+\exp[-\lambda(p-\theta_{\text{syn}})]\}^{-1}\,$, which acts as a continuous mechanism for the activation and deactivation of the chemical synapses. The coupling strength associated to this type of synapses is $g_{ch}$. For the chosen set of parameters, $|p_i|<2$ and thus $(p_i-V_{\text{syn}})$ is always negative, meaning that the chemical coupling is excitatory if $V_{\text{syn}}=2$. The other parameters are $\theta_{\text{syn}}=-0.25$ and $\lambda=10$ following Refs. \cite{Baptistaetal2010,Antonopoulosetal2015}. For simplicity, we assume that both types of synapses are bidirectional.

Additionally to Eqs.~\eqref{HR_model_Nneurons}, we introduce the instantaneous angular frequency of the $i$-th neuron as \cite{Pereiraetal2007},
\begin{equation}
\label{HR_model_phase}
\dot{\phi}_i=\frac{\dot{q}_i p_i-\dot{p}_i q_i}{p_i^2+q_i^2},
\end{equation}
where $\phi_i$ is the phase defined by the fast variables, $p_i$ and $q_i$ of the $i$-th neuron; $i=1,\ldots,N$.

In the following, we analyze the dynamical behavior of the system of Eqs.~\eqref{HR_model_Nneurons} by means of numerical simulations for a network of six small-world communities, detected in the \celegans~neural network. We also link the emergent dynamics to the coaction of the two types of synapses. The large-scale interaction between communities is illustrated in Fig.~\ref{fig1}(b), where the communities are represented by circles of sizes proportional to their number of nodes. The width of each arrow is proportional to the number of chemical synapses between two communities and the size of the arrowhead is proportional to the relative influence of one community to the other, that is, the number of the chemical synapses divided by the mean degree of the electrical synapses of the target community. A more detailed analysis reveals that only $20 \%$ of the neurons are exclusively connected within their community (Fig.~\ref{fig1}(c)). The rest of neurons are linked with neurons in other communities through 
chemical synapses. As in other brain networks \cite{Zamora2011, Klimm2014}, a trend is appreciated for which, sparsely connected neurons have low participation while the hubs span their links among many communities (Fig.~\ref{fig1}(d)). In this case, all hubs of the network are condensed into community 4 and spread links among five or six communities. In conclusion, community 4 is not only found to be the largest but also the most influential one in terms of its inter-modular connectivity.

In the Supplementary Information we present the same analysis using modular networks with 277 neurons grouped into six communities, which share some common features with those of the \celegans~neural network.

\paragraph*{\bf Modular Structure of the \celegans~Neural Network.}

We identified the communities of the \celegans~neural network employing the walktrap method \cite{Ponsetal2005}  of the igraph software, using six steps, following Ref. \cite{Antonopoulosetal2015}. The algorithm detects communities through a series of short random walks, based on the idea that the nodes encountered on any given random walk are more likely to be within a community. The algorithm initially treats all nodes as communities of their own, then merges them into larger communities, and these into even larger ones, and so on. Essentially, it tries to find densely connected subgraphs (\ie{} communities) in a graph via random walks. The idea is that short random walks tend to stay in the same community. Following this approach, we have been able to identify six communities in the \celegans~neural network.

\paragraph*{\bf Contribution of Nodes on the Mesoscale.}

Given a modular network, its nodes can play different roles according to their inter- and intra-modular connections \cite{Guimera2005a,Guimera2005b}. For example, some nodes may only connect to nodes in a particular community. Such nodes, often referred to as \emph{peripheral}, form the core of the community. Other nodes might be well-connected within a module but extend connections to other modules. These allow for the communities to link with each other. In the other extreme case we might find nodes which are equally connected across all modules and therefore, are difficult to be classified within modules. Such nodes, often called \emph{kinless nodes}, may however allow for the integration of information, since they have access to the information in all modules. Here, we characterize the roles of the nodes based on a recently introduced appropriate framework \cite{Klimm2014}.
\\

\noindent\emph{Hubness:}
In order to characterize the hubness of a node so that its hubness value is comparable across networks of different sizes and densities, the degrees of a node has to be evaluated under a common reference or statistical frame. We define the hubness index $h_i$ of a node $i$ in a network of size $N$ and density $d$ as the relative difference of the node's degree $k_i$ with respect to the degree distribution of an equivalent random graph of the same size and number of nodes,
\begin{equation}
h_i = \frac{k_i - \left< k \right>_{R}}{\sigma_{R}} = \frac{k_i - (N-1) d}{\sqrt{(N-1) \, d \, (1-d)}},
\label{eq:HubnessIndex}
\end{equation}
where $\left< k  \right>_R$ is the mean degree of the equivalent random graph, and $\sigma_R$ the standard deviation of its degree distribution. The hubness is negative for nodes with degree $k_i <  \left< k \right>_{R}$, allowing also for the identification of outliers that are significantly less connected than expected from randomness. 
\\

\noindent\emph{Participation Index:}
Given a partition $\mathcal{C}$ with $M$ disjoint communities, the aim is to quantify how distributed a node is among them. The contribution of a node to each community depends on the community size and therefore,  its participation is characterized in terms of the likelihood the node has to belong to a given community. A \emph{participation vector} {\boldmath $F_i$} with elements $F_{im}$ represent the probability that node $i$ belongs to community $C_m$, where $m = 1, 2, \ldots, M$. This probability is given by $F_{im} = \frac{k_{im}}{N_m}$, where $N_m$ is the size of the community. Since we are only interested in the relative differences, the participation vectors are normalized such that $\sum_{m=1}^M F_{im} = 1$. The vector of a node devoting all its links to the second community of a network with $M=6$ communities is {\boldmath $F_i$}~$= (0,1,0,0,0,0)$ and, for a node whose links are all equally likely distributed among the six communities, {\boldmath $F_i$}~$=(1/6,1/6,1/6,1/6,1/6,1/6)$. When reducing 
this information into an index that quantifies how distributed the links of a node are among all communities, for consistency with previous definitions \cite{Guimera2005a}, we set $f_i = 0$ if the node devotes all its links to a single community and $f_i = 1$ if its links are equally likely distributed among \emph{all} modules. Therefore, we evaluate the standard deviation $\sigma(\mathbf{F_i})$ of the elements of the participation vector {\boldmath $F_i$} and, define the participation index as
\begin{equation}
f_i \: = \: 1 - \frac{\sigma(\mathbf{F_i})}{\sigma_{max}(M)} \: = \: 1 - \frac{M}{\sqrt{M-1}} \; \sigma(\mathbf{F_i}).
\label{eq:pindex}
\end{equation}
The normalization factor accounts for the fact that the standard deviation of an $M$-dimensional vector with all elements equal to zero but one is $\sigma_{max}(M) = \sqrt{M-1} \, / \, M$.

\paragraph*{\bf Estimating Synchronization, Metastability and Chimera-like States.}

In this paper we use the order parameter $\rho$ to account for the synchronization level of the neural activity of the considered networks and of their communities \cite{Gardnesetal2010}. It is originated from the theory of measures of dynamical coherence of a population of $N$ Kuramoto phase oscillators \cite{KUR02a} and can be computed by a complex number $z(t)$ defined as
\begin{equation}
\label{z_t}
z(t)=\rho(t)e^{\mathrm{i}\Phi(t)}=\frac{1}{N}\sum_{j=1}^{N}e^{\mathrm{i} \phi_j(t)}.
\end{equation}
By taking the modulus $\rho(t)$ of $z(t)$, one can measure the phase coherence of the $N$ neurons of the network, and by $\Phi(t)$ their average phase. In this context, $\phi_i$ is the phase variable of the $i$-th neuron given by Eq.~\eqref{HR_model_phase}. Actually, one averages $\rho(t)$ in time to obtain the order parameter $\rho=\langle\rho(t)\rangle_t$, namely the tendency of $\rho(t)$ in time. A value of $\rho=1$ corresponds to complete synchronous activity, whereas $\rho=0$ to complete desynchronization. We use Eq.~\eqref{z_t}, adapted accordingly, wherever we need to compute the synchronization level of neural networks or communities. In particular, in the case of the whole neural network, $N$ is the total number of neurons, whereas in the case of communities, $N$ represents the number of neurons of the particular community.

The previous paragraph focused on the synchronization of each community as described by the time average of the order parameter. However, if one wants to look at the instantaneous behavior of each community and of the whole network, one will see that there are alternating time intervals of synchronization and incoherence for each community, also seen in the incoherent regions of Fig.~\ref{fig2}. This is due to the chaotic bursting behavior of the \hr~model which involves quiescent periods, where neurons easily synchronize, followed by fast spiking intervals where neurons tend to desynchronize because of the corresponding faster time scales. 

Complex dynamical systems do not necessarily settle into stationary synchronized states. Instead, they can be metastable in time, meaning that they can stay in the vicinity of one stable state for a certain time interval and then, spontaneously move towards another. An even more interesting feature of many complex systems is undoubtedly the coexistence of different, often contradictory states. In terms of synchronization, this is illustrated through the so-called {\it chimera} states \cite{KUR02a,ABR04}, where one population of oscillators synchronizes whereas other populations of identical oscillators are desynchronized.

In order to quantify the metastability and chimera-likeness of the observed dynamics, we employ two measures first introduced by M. Shanahan in Ref. \cite{SHA10}. In particular, the level of metastability can be calculated from the so-called metastability index $\lambda$, given by the expression
\begin{equation}
\label{lambda}
\lambda = {\langle \sigma_{\textbf{met}} \rangle}_{C_m},
\end{equation}
where,
\begin{equation}
\label{sigma_lambda}
\sigma_{\textbf{met}}(m)=\frac{1}{T-1}\sum_{t=1}^T(\rho_m(t)-{\langle \rho_m \rangle}_T)^2. 
\end{equation}
In Eq.~\eqref{lambda}, $C_m$ is the set of all $M$ communities and $m=1,2,\ldots,M$. The order parameter $\rho_m(t)$ of each community $m$ is sampled at discrete times $t \in {1,\ldots,T}$. For a given community, the variance $\sigma_{\textbf{met}}(m)$ of $\rho_m(t)$ over all time steps, gives an indication of how much the synchrony in this community fluctuates in time. Averaging over all $M$ communities, one obtains an index of the metastability present in the entire network.

Similarly, the so-called chimera-like index $\chi$ \cite{SHA10} is given by
\begin{equation}
\label{chi}
\chi={\langle \sigma_{\textbf{chi}} \rangle}_T,
\end{equation}
where
\begin{equation}
\label{sigma_chi}
\sigma_{\textbf{chi}}(t)=\frac{1}{M-1}\sum_{m=1}^M(\rho_m(t)-{\langle \rho(t) \rangle}_M)^2. 
\end{equation}
In the above expression, $\sigma_{\textbf{chi}}(t)$ is an instantaneous quantity that gives the variance of $p_m(t)$ over all $M$ communities at a given time $t$. The time average of this quantity indicates how chimera-like a certain state is.

\providecommand{\noopsort}[1]{}\providecommand{\singleletter}[1]{#1}%

\section*{Author contributions statement}
J.~H., N.~E.~K. and C.~G.~A. conceived and designed 
the study and performed the numerical simulations. 
G.~Z.-L. and A.~D.~-G. analyzed the roles of the nodes. 
J.~H., N.~E.~K., C.~G.~A., G.~Z.-L. and A.~D.~-G.  
carried out the analysis and wrote the article.
%

\section*{Acknowledgments}
This study was supported by the J.~S.~Latsis Public Benefit Foundation in Greece. J.~H. acknowledges support by the EU/FP7-REGPOT-2012-2013-1 under grant agreement n316165. N.~E.~K. and A.~D.~-G. acknowledges support by the LASAGNE (Contract No.318132) and MULTIPLEX (Contract No.317532) EU projects as well as by the Generalitat de Catalunya (2014SGR608) and Spanish MINECO (FIS2012-38266). G.~Z.~-L. received support from the European Union Seventh Framework Programme FP7/2007-2013 under grant agreement number PIEF-GA-2012-331800.

\section*{Additional information}
The authors declare no competing financial interest. Correspondence or request for material should be addressed to J.~H., N.~E.~K., C.~G.~A, G.~Z.~-L. or A.~D.~-G..

\section*{Figures Legends}

\end{document}